\pdfoutput=1
\documentclass[10pt,conference]{IEEEtran}
\IEEEoverridecommandlockouts
\usepackage{url}
\usepackage{xspace}
\usepackage{amsmath,amssymb,amsfonts}
\usepackage{algorithmic}
\usepackage{graphicx}
\usepackage{textcomp}
\usepackage{xcolor}
\usepackage{verbatim}
\usepackage{multirow}
\usepackage{subfig}
\usepackage{booktabs} 
\usepackage{enumitem}
\usepackage{tabularx}
\usepackage{bm}

\usepackage{cleveref}
\crefformat{section}{\S#2#1#3}
\crefname{figure}{Figure}{Figures}
\crefname{table}{Table}{Tables}

\usepackage{todonotes}

\newcommand{\HIDDEN}[1]{\todo[disable]{#1}}

\definecolor{gamboge}{rgb}{0.89, 0.61, 0.06}
\definecolor{applegreen}{rgb}{0.55, 0.71, 0.0}
\definecolor{azure}{rgb}{0.0, 0.5, 1.0}

\newcommand{\obs}[1]{#1}
\newcommand{\interp}[1]{#1}

\newcommand{\ie}{\textit{i.e.,}\xspace}
\newcommand{\eg}{\textit{e.g.,}\xspace}
\newcommand{\etal}{\textit{et al.}\xspace}

\makeatletter
\newcommand{\mysubsection}[1]{\vspace{1.5mm}\noindent\textbf{#1\@addpunct{.}}}
\newcommand{\mysubsubsection}[1]{\vspace{1.5mm}\noindent\textit{#1\@addpunct{.}}}
\newcommand{\mysubsubsubsection}[1]{\noindent{#1\@addpunct{.}}}
\makeatother

\newenvironment{smalldescriptionrq}{
	 \setlength{\topsep}{0pt}
	 \setlength{\partopsep}{0pt}
	 \setlength{\parskip}{0pt}
	 \begin{description}[style=unboxed]
	 \setlength{\leftmargin}{-1in}
	 \setlength{\parsep}{0pt}
	 \setlength{\parskip}{0pt}
	 \setlength{\itemsep}{0pt}
	 }
	 {\end{description}}

\def\BibTeX{{\rm B\kern-.05em{\sc i\kern-.025em b}\kern-.08em
    T\kern-.1667em\lower.7ex\hbox{E}\kern-.125emX}}
\begin{document}

\title{On the Strategies to Improve Continuous Integration: Reducing Time to Feedback and Computational Cost}
\title{What helped, and what did not? An Evaluation of the Strategies to Improve Continuous Integration}

\author{
\IEEEauthorblockN{Xianhao Jin}
\IEEEauthorblockA{\textit{Computer Science} \\
\textit{Virginia Tech}\\
xianhao8@vt.edu}
\and
\IEEEauthorblockN{Francisco Servant}
\IEEEauthorblockA{\textit{Computer Science} \\
\textit{Virginia Tech}\\
fservant@vt.edu}
}

\maketitle
\thispagestyle{plain}
\pagestyle{plain}

\begin{abstract}
Continuous integration (CI) is a widely used practice in modern software engineering.
Unfortunately, it is also an expensive practice — Google and Mozilla estimate their CI systems in millions of dollars.
There are a number of techniques and tools designed to or having the potential to save the cost of CI or expand its benefit - reducing time to feedback.
However, their benefits in some dimensions may also result in drawbacks in others.
They may also be beneficial in other scenarios where they are not designed to help.
In this paper, we perform the first exhaustive comparison of techniques to improve CI, evaluating 14 variants of 10 techniques using selection and prioritization strategies on build and test granularity.
We evaluate their strengths and weaknesses with 10 different cost and time-to-feedback saving metrics on 100 real-world projects.
We analyze the results of all techniques to understand the design decisions that helped different dimensions of benefit.
We also synthesized those results to lay out a series of recommendations for the development of future research techniques to advance this area.
\end{abstract}


\section{Introduction}
Continuous Integration (CI) is a software development practice by which developers integrate code into a shared repository several times a day \cite{fowler2006continuous}.
However, CI gains adoption in practice, difficulties \eg \cite{pinto2017inadequate} and pain points \eg \cite{widder2019conceptual} have been discovered about it.
As software companies adopt CI, they execute builds for many of projects, and they do so very frequently.
As workload increases, two main problems appear:
(1) the time to receive feedback from the build process increases, as software builds often outnumber the available computational resources --- having to wait in build queues,
and
(2) the computational cost of running builds also becomes very high.
Previous studies \eg \cite{memon2017taming} have highlighted the long time that developers have to wait to receive feedback about their builds.
For example, at Google, developers must wait 45 minutes to 9 hours to receive testing results \cite{liang2018redefining}.
Even just the dependency-retrieval step of CI can take up to an hour per build \cite{celik2016build}.
Regarding the high cost of running builds, that is also highlighted in other studies \cite{herzig2015art, hilton2017trade, hilton2016usage, pinto2017inadequate, widder2019conceptual}.
The cost of CI reaches millions of dollars, \eg at Google \cite{hilton2016usage} and Microsoft \cite{herzig2015art}.
While other problems exist for CI, we focus on these two because they are the ones that most existing techniques have focused on addressing.
They are also interrelated, since cost-reduction techniques may also reduce time-to-feedback --- \eg skipping some tests may cause other tests to fail earlier.

Multiple techniques have been proposed to improve CI.
Most of them have the goal of reducing either its \textbf{time-to-feedback} or its \textbf{computational cost}.
All such techniques consider the observation of build failures to be more valuable than build passes, because failures provide actionable feedback, \ie they point to a problem that needs to be addressed.
\textbf{Time-to-feedback-reduction} techniques aim to observe \textbf{failures earlier} --- by \textbf{prioritizing} failing executions over passing ones.
These techniques may operate in two different levels of granularity, by prioritizing: test executions \eg \cite{elbaum2014techniques}, or build executions \eg \cite{liang2018redefining}.
\textbf{Computational-cost-reduction} techniques aim to observe \textbf{failures only} --- by \textbf{selectively executing} failing builds only, saving the cost of executing passing ones.
They also may operate at two different levels of granularity, selecting: test executions \eg \cite{Machalica2019predictive}, or build executions \eg \cite{abdalkareem2019commits}.

To the extent of our knowledge, the existing techniques to improve CI have been evaluated under different settings, making it hard to compare them.
Previous studies used different software projects, different metrics, and rarely compared one technique to another.
However, we expect that different choices of goal, granularity, and technique design will bring different trade-offs.
For example, cost-reduction techniques at build-granularity may be more \emph{risky} than a test-granularity one, \ie it may save more cost when it skips all the tests in a build, but it may also make more mistakes if it skips many failing tests in a build.
However, the opposite may be true, if test-granularity cost-reduction techniques also skip a large ratio of full builds (\ie all the tests in the build).
On another example, test-selection techniques may be a good alternative to test-prioritization techniques that also saves cost as an added benefit, or they may instead delay the observation of test failures if they mispredict too many of them.
To the best of our knowledge, how these trade-offs manifest in practice is still mostly unknown.
Empirically understanding these trade-offs will have valuable practical implications for the design of future techniques and for practitioners adopting them.

In this paper, we perform the first evaluation of the existing strategies to improve CI.
We aim to understand the trade-offs between these techniques
for three dimensions: 
(D1) computational-cost reduction,
(D2) missed failure observation, and
(D3) early feedback.

For this goal, we performed a large-scale evaluation.
We replicated and evaluated all the existing 10 CI-improving techniques from the research literature, 
representing the two goals (time-to-feedback and computational-cost reduction) and the two levels of granularity (build-level and test-level) for which such techniques have been proposed.
We evaluated these techniques under the same settings, using the state-of-the-art dataset of continuous-integration data: TravisTorrent \cite{msr17challenge}.
To be able to study all techniques, we extended TravisTorrent in multiple ways, mining additional Travis logs, Github commits, and building dependency graphs for all our studied projects.
Finally, we measured the effectiveness of all techniques with 10 metrics in 3 dimensions.
We included every metric that any previous evaluation of our studied techniques used (7), refitted 2 others and designed an additional one.

We analyzed the results obtained by all techniques on all metrics across all 3 dimensions, and we synthesized our observations, to understand which design decisions helped and which ones did not for each dimension.
Finally, we further reflect on our results to provide a wide set of recommendations for the design of future techniques in this research area.

The main contributions of this paper are:
(1) the first comprehensive evaluation of CI-improving techniques;
(2) a collection of metrics to measure the performance of CI-improving techniques over various dimensions;
(3) an extended Travis Torrent dataset with: detailed test and commit, and dependencies information;
(4) the replication of 14 variants of 10 CI-improving techniques;
(5) evidence for researchers to design future CI-improving techniques.

\section{Approaches to Improve Continuous Integration}
\label{sec:taxonomy}

\begin{figure*}
\centering
\includegraphics[width=0.8\linewidth]{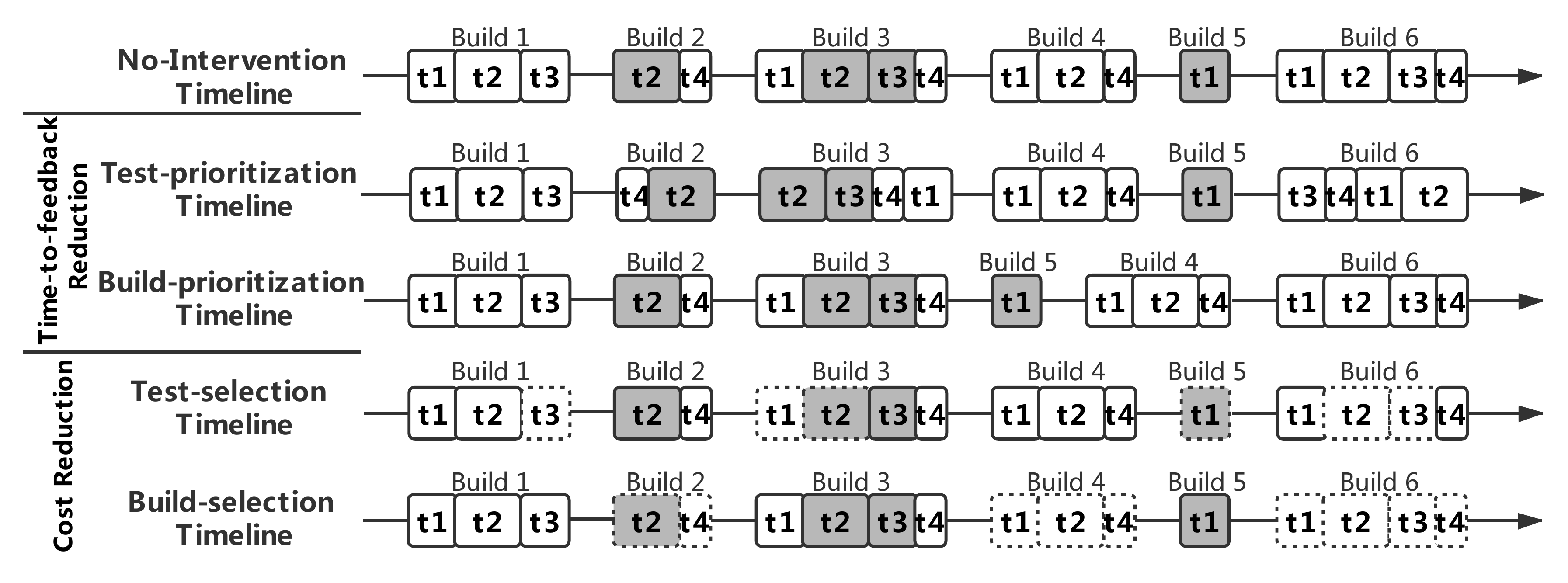}
\caption{
Example timeline. 
Failing tests in gray. 
Build-selection runs builds fully when it predicts a failing build. 
Test-selection runs builds partially (for tests that would fail). 
Build-prioritization changes the build sequence. 
Test-prioritization changes the test sequence within a build.
}
\label{fig:builds}
\end{figure*}

We summarize technique families in \cref{tab:techniques} and discuss each technique in detail in \cref{sec: Techniques}.
Figure~\ref{fig:builds} depicts a non-interventional example timeline of builds, a timeline in which a build-selection technique is applied, a timeline produced by build-prioritization technique, a timeline where a test-selection technique is applied, and a timeline with applying a test-prioritization approach.
The example timeline shows a chronological numbered sequence of builds in CI.
Each build is made up of at least one test.
We depict each test suite as a rectangle with a test number (e.g., t1).
Failing tests are then highlighted in gray.
The length of the rectangle refers to the time duration for the test to be executed.
We depict skipped tests with a dashed rectangle.
In the most ideal cost-saving scenario, all of the passing tests would be skipped and all of the failing tests would be observed as soon as possible.

\subsection{Computational-cost Reduction}

\subsubsection{Test-level granularity}
Test-selection techniques \cite{gligoric2015practical,herzig2015art, Machalica2019predictive, memon2017taming, shi2017optimizing, zhang2018hybrid, zhu2019framework} aim at automatically detect and label tests that are not going to fail.
These test-level approaches collect information from test history and project dependency along with the current commit and use some heuristic models to detect failing tests and skip the others.
Figure~\ref{fig:builds} also illustrates how this type of techniques works in the simulation timeline.
After a test-selection approach is activated, it selects a subset of tests (e.g., t2 in build \#2, t4 in build \#4) that it predicts to have a possibility to fail and decides to skip the others (e.g., t3 in build \#1, t1 in build \#5).
For those tests that are not selected in the timeline and get skipped, we depict them as dashed rectangles.
In this paper, we consider it can skip some builds when it selects no test in those builds.

\subsubsection{Build-level granularity}
Build-selection techniques \cite{abdalkareem_tse2020, abdalkareem2019commits, hassan2017change, jin2020_icse, ni2017cost} aim at automatically detect and label commits and builds that can be CI skipped.
Some approaches  \cite{hassan2017change, jin2020_icse, ni2017cost} try to detect failing builds and skip those passing builds to achieve cost-saving.
Others \cite{abdalkareem_tse2020, abdalkareem2019commits} aim at identifying commits that can be CI skipped.
\cref{fig:builds} illustrates how they work in the simulation timeline.
As a build-level technique, when build-selection approach decides to skip a build (e.g., build \#2, \#4, \#6), normally it skips all of the tests in that build.
The inner test sequence is not changed and all of tests are run in an executed build.

\subsection{Time-to-feedback Reduction}

\subsubsection{Test-level granularity}
Test-prioritization techniques \cite{elbaum2014techniques, luo2018assessing, marijan2013test, mostafa2017perfranker, thomas2014static} try to give high priority to tests that are predicted to be failed so that developers could be informed in a shorter time.
This family of approaches normally rearrange the execution order of tests within a build to make predicted-to-fail tests run earlier by analyzing information such as test failing history and test context.
Figure~\ref{fig:builds} depicts an example of how this type of techniques works in the simulation timeline.
With a test-level approach being activated, the CI system gives different tests different priorities and firstly executes those tests with a higher priority (e.g., t4 in build \#2, t2 in build \#3) as well as delays low-priority tests (e.g., t1 in build \#3, t2 in build \#6).
The sequence of test executions in this timeline gets rearranged and the start-time for tests that are more likely to fail move ahead in time.
Also, all tests are executed at last.


\subsubsection{Build-level granularity}
Build-prioritization techniques 
\cite{liang2018redefining} aim at automatically
prioritizes commits that are waiting for being executed.
They favor builds with a larger percentage of test suites that have been found to fail recently and builds including test suites that have not been executed recently as an alternative path.
Figure~\ref{fig:builds} also shows how this family of techniques works in the simulation timeline.
Build-prioritization techniques will only be activated when there is a collision of builds (i.e., there are multiple builds waiting to occupy the limited resource).
The technique is build-level so it will not change the inner order of the test executions and it will normally change the sequence of tests across builds when the approach is activated (e.g., build \#4, \#5).
None of tests become dashed in this timeline because they all eventually execute.



\section{Research Method}
In this paper, we replicated and evaluated 14 variants of 10 CI-improving techniques, covering their two goals (time-to-feedback and computational-cost reduction) and their two levels of granularity (build-level and test-level) with 1 perfect technique for the ideal timeline.
We evaluate them over 100 software projects in TravisTorrent, which we extended to be able to run all such kinds of techniques.

Our goal is to understand the trade-offs between existing CI-improving techniques, and between the metrics that have been used to evaluate them.
We perform 2 empirical studies to analyze these trade-offs for the following 3 dimensions of CI-improving techniques, using 10 metrics.
We only include selection techniques in Empirical Study 1 since prioritization techniques have no power in cost saving by nature.
We involve selection and prioritization techniques in Empirical Study 2 because both of them can have an impact on fault detection, \eg wrongly-skipped failing builds by selection approaches can cause delay in fault detection.

\begin{smalldescriptionrq}
\item[Empirical Study 1: Cost Saving]
\item[\ \ D1:] \textbf{Computational-cost Reduction}
 \item[\ \ D2:] \textbf{Missed Failure Observation}
\item[Empirical Study 2: Time-to-feedback Reduction]
 \item[\ \ D3:] \textbf{Early Feedback}
\end{smalldescriptionrq}


For each dimension, we study:
\begin{smalldescriptionrq}
\item[RQ1:] \textbf{What design decisions helped this dimension?}
\item[RQ2:] \textbf{What design decisions did not help this dimension?}
\end{smalldescriptionrq}
\subsection{Data Set}
We perform our study over the Travis Torrent dataset \cite{beller2017oops}, which includes 1,359 projects (402 Java projects and 898 Ruby projects) with data for 2,640,825 build instances. 
We remove ``toy projects'' from the data set by studying those that are more than one year old, and that have at least 200 builds and at least 1000 lines of source code, which is a criteria applied in multiple other works \cite{ni2017cost,islam2017insights}. 
To be able to evaluate test-granularity techniques, we also filter out those projects whose build logs do not contain any test information.
We focused our study on builds with passing or failing result, rather than error or canceled --- since they can be exceptions or may happen during the initialization and get aborted immediately before the real build starts.
Besides, in Travis a single push or pull-request can trigger a build with multiple jobs, and each job corresponds to a configuration of the building step.
We did a preliminary investigation of these builds and found that these jobs with the same build identifier normally share the same build result and build duration.
Thus, as many existing papers have done \cite{gallaba2018noise,rebouccas2017does,jain2019brief}, we considered these jobs as a single build.
After this filtering process, we obtained 82,427 builds from 100 projects (13,464 failing builds).

To be able to execute all our studied techniques, we extended the information in TravisTorrent of these 100 projects in multiple ways.
First of all, we needed to know the duration of each individual test for the comparison and replication.
Also, to replicate some techniques, \eg \cite{herzig2015art,elbaum2014techniques}, we needed to capture the historical failure ratio for each individual test.
To obtain these information, we built scripts to download the raw build logs from Travis and parse them to extract all of the information about test executions, such as test name, duration and outcome.
Some techniques, \eg \cite{Machalica2019predictive, abdalkareem2019commits}, require additional information that TravisTorrent does not provide for builds, such as the content of commit messages, changed source lines and changed file names.
For that, we also mined additional information about commits in the projects' code repositories through Github.
Then, we matched each test with its corresponding test file in the project. 
Finally, to be able to run other techniques, \eg \cite{gligoric2015practical,Machalica2019predictive}, we built a dependency graph for the source code of each project using a static code analysis tool (Scitool Understand \cite{SciTool}) to determine the paths between the source files and test files.

\subsection{Evaluation Process}
\label{evaluation_processs}
We evaluate the techniques in a real-world scenario, to understand as best as possible the behavior that the techniques would show in practice.
We take two measures for that.

First, we respect the original chronological order of build and test operations when training techniques.
We achieve that by using an 11-fold, chronological variant of cross-validation.
For each project, we split its chronological timeline into 11 folds.
We use the first chronological fold only for testing, and we iteratively test the other 10 folds.
For each testing fold, we train on all the folds that precede it chronologically.
This approach has been used in previous works \eg \cite{bettenburg2008duplicate,servant12icse} to avoid training with information that would not be available in practice, \ie it happens in the future.

We follow this approach for all the techniques based on machine learning, \eg \cite{Machalica2019predictive}.
For techniques that do not require training, \eg \cite{abdalkareem2019commits}, we simply execute them over the same last 10 folds.
For techniques that train on data from other projects, \ie for cross-project technique variants, we also executed them over the same last-10-fold timeline --- and we divided them into 10 \emph{project} folds to do cross-project cross-validation, \ie for each project, the technique is trained on 90 other projects and tested on its last 10 fold data.

Second, we respect the real-world availability of information.
That is, for selection-based techniques, when a build or test is skipped, the technique will not know its outcome.
For techniques that rely on the last build or test outcome \eg \cite{hassan2017automatic}, we only inform them of the outcome of the last \emph{executed} build or test.
Additionally, when builds are skipped, we accumulate their code changes into the subsequent build.


\begin{table}
\caption{
Studied Techniques.
}
\label{tab:techniques}
  \label{tab:overview}
\small
\begin{tabular}{|m{1cm}|p{1.6cm}|p{1.6cm}|p{2.6cm}|}
\hline
\multicolumn{1}{|c|}{\textbf{Goal}} & \multicolumn{1}{c|}{\textbf{Approach}} & \multicolumn{1}{c|}{\textbf{Granularity}} & \multicolumn{1}{c|}{\textbf{Studied Technique}} \\ \hline
\multirow{4}{1cm}{Time to Feed\-back}   & \multirow{4}{*}{Prioritization} & \multirow{3}{*}{Test}  & PT\_Marijan13 \cite{marijan2013test}              \\ \cline{4-4} 
                                    &                                 &                        & PT\_Elbaum14 \cite{elbaum2014techniques}               \\ \cline{4-4} 
                                    &                                 &                        & PT\_Thomas14 \cite{thomas2014static}               \\ \cline{3-4} 
                                    &                                 & Build                  & PB\_Liang18 \cite{liang2018cost}               \\ \hline
\multirow{6}{1cm}{\shortstack[l]{Comput-\\ational \\ \\ Cost}} & \multirow{6}{*}{Selection}      & \multirow{3}{*}{Test}  & ST\_Gligoric15 \cite{gligoric2015practical}            \\ \cline{4-4} 
                                    &                                 &                        & ST\_Herzig15 \cite{herzig2015art}               \\ \cline{4-4} 
                                    &                                 &                        & ST\_Mach19 \cite{Machalica2019predictive}                \\ \cline{3-4} 
                                    &                                 & \multirow{3}{*}{Build} & SB\_Hassan17 \cite{hassan2017change}               \\ \cline{4-4} 
                                    &                                 &                        & SB\_Abd19 \cite{abdalkareem2019commits}                 \\ \cline{4-4} 
                                    &                                 &                        & SB\_Jin20 \cite{jin2020_icse}                 \\ \hline
\end{tabular}
\vspace{-.2in}
\end{table}

\subsection{Replicated Techniques}
\label{sec: Techniques}
We replicated and studied all the techniques that have been proposed to improve CI by reducing the time to feedback or reducing its cost.
In addition to these, there are other techniques that were proposed before CI and that could also be applied for these two goals: test prioritization techniques, and test selection techniques.
Therefore, we also replicated and studied a state-of-the-art technique in each of these two categories that were not originally proposed for CI.
We summarize all our studied techniques in Table~\ref{tab:overview}.

In total, we studied 10 techniques, across two goals (reducing time to feedback and cost) and two granularities (test and build levels).
Since we also studied multiple variants of some techniques, our evaluation included 14 total technique variants.
To provide a reference point, we also studied a perfect technique:
\textit{Perfect Technique}. 
It achieves the goal of each metric perfectly --- it predicts which tests or builds will fail with 100\% accuracy, prioritizing or selecting them perfectly.

We include the detailed description for each technique in \cref{sec:tech1} and \cref{sec:tech2}.

\section{Empirical Study 1: Cost Saving}

\subsection{Studied Techniques}
\label{sec:tech1}


\subsubsection{Test-selection Techniques}
We replicated all the test-selection techniques that were proposed for improving CI: ST\_Mach19 \cite{Machalica2019predictive} and ST\_Herzig15 \cite{herzig2015art}.
To provide even more context for our study, we also evaluate a state-of-the-art test-selection technique: ST\_Gligoric15 \cite{gligoric2015practical} --- since test-selection techniques have also been proposed outside the context of CI, \eg \cite{zhang2018hybrid,gligoric2015practical,yoo2012regression,yoo2007pareto,rothermel1997safe,rothermel1996analyzing}.


\noindent\textbf{ST\_Gligoric15 \cite{gligoric2015practical}} skips tests that cannot reach the changed files, by tracking dynamic dependencies of tests on files.
A test can be skipped in the new revision if none of its dependent files changed.
The rationale is that tests that cannot reach changed files cannot detect faults in them.

\noindent\textbf{ST\_Herzig15 \cite{herzig2015art}} is based on a cost model, which dynamically skips tests when the expected cost of running the test exceeds the expected cost of removing it, considering both the machine cost and human inspection cost \cite{bell2018deflaker, herzig2015empirically}.
This technique tends to skip tests that mostly passed in the past or that have long runtime.


\noindent\textbf{ST\_Mach19 \cite{Machalica2019predictive}} proposes a Machine Learning algorithm with combined features of commit changes and test historical information.
We studied two variants of it: one is trained in the past builds within the same project in which it is applied (\emph{ST\_Mach19\_W}), and the other is trained in the builds of different software projects than the one in which it will be applied (\emph{ST\_Mach19\_C}).
It uses the following features: ﬁle extensions, change history, failure rates, project name, number of tests and minimal distance.

\subsubsection{Build-selection Techniques} 
We then replicated all build-selection techniques that jave been proposed for improving CI: SB\_Abd19 \cite{abdalkareem2019commits}, and SB\_Jin20 \cite{jin2020_icse}.
To provide even more context for our study, we also replicated a state-of-the-art build-prediction technique: SB\_Hassan17 \cite{hassan2017change}.

\noindent\textbf{SB\_Hassan17 \cite{hassan2017change}} predicts every build's outcome based on the information from last build.
Builds can be skipped when they are predicted to pass.
In our study, information from the previous build is blinded if the build does not get executed.
We study two variants of this technique (\emph{SB\_Hassan17\_W} and \emph{SB\_Hassan17\_C}) as we did for \emph{ST\_Mach19}.
\noindent

\noindent\textbf{SB\_Abd19 \cite{abdalkareem2019commits}} uses a rule-based approach to skip commits that only have \emph{safe} changes, \eg changes on configuration or document files.
This technique is expected to capture most failing builds since it only skips builds considered safe to skip.

\noindent\textbf{SB\_Jin20 \cite{jin2020_icse}} aims at saving CI cost by skipping passing builds.
Their strategy is to capture the first failing build in a subsequence of failing builds and continuously build until a passing build appears.
We replicated this technique under the configuration that provided the optimal effectiveness \cite{jin2020_icse}.
We studied three variants of this technique: \emph{SB\_Jin20\_W} \& \emph{SB\_Jin20\_C} as we did previously, and also a rule-of-thumb variant (SB\_Jin20\_S) that skips builds with $<4$ changed files.

\subsection{D1: Computational-cost Reduction}
\label{sec:d1}

We studied four metrics for D1.
We plot the result of each metric in a box plot where each box represents the distribution of values for all the studied projects.

\subsubsection{Studied Metrics}
\label{sec:rq1_metric}

\noindent\textbf{Build time saved}
measures the proportion of total build time that is skipped among all build time per project.
It was covered in SB\_Abd19 \cite{abdalkareem2019commits}.

\noindent\textbf{Test time saved}
measures the same as the previous metric but in terms of test time.
The previous work ST\_Gligoric15 \cite{gligoric2015practical} used this metric in its evaluation.
It shows how much time applying a technique could save during the phase of test executions.

\noindent\textbf{Builds number saved}
measures the proportion
of builds that are saved among all builds.
It was studied by SB\_Abd19 \cite{abdalkareem2019commits} and SB\_Jin20 \cite{jin2020_icse}. 
It represents how many resources could be saved as the number of builds.

\noindent\textbf{Tests number saved}
measures the same as the previous metric but in term of tests.
Previous papers \cite{gligoric2015practical,herzig2015art} studied this metric.
It represents how many resources could be saved during test executions.


\subsubsection{Analysis of Results}
\label{sec:d1rq1}


	
	
\mysubsection{Comparing Metrics}
When we compare the techniques' 
test number vs. test time saved,
\obs{most of them saved a very similar ratio of test time than ratio of tests (except ST\_Herzig15)}.

When comparing 
build number vs. build time,
\obs{build-granularity techniques saved a very similar ratio of build time as of builds}.
Also, \obs{test-granularity techniques saved a larger ratio of build time than of builds}.
This means that \interp{test-granularity techniques save build time when they skip builds partially --- when they skipped some of their tests}.

When comparing 
test number vs. build number,
\obs{build-granularity techniques saved a very similar ratio of builds and tests}.
Also, \obs{test-granularity techniques saved a much lower ratio of builds than of tests} --- some dramatically so (ST\_Herzig15 and ST\_Mach19\_C).
This means that \interp{test-granularity techniques saved a low ratio of full builds}.

When comparing 
test time vs. build time,
\obs{build-granularity techniques saved very similar ratios of test time and build time}.
Also, \obs{test-granularity techniques saved a much lower ratio of build time than of test time}.
This observation extends our earlier one: every build that these techniques did not skip fully, and thus did not skip its build-preparation time, reduced their ability to save build time to an important extent.

\mysubsection{Comparing Granularities}
By comparing test vs. build-granularity techniques,
\obs{build-granularity techniques generally saved higher build-time cost} --- except for SB\_Abd19.
Build-granularity techniques have the advantage of skipping both test-execution and build-preparation time, while test-granularity techniques have the advantage of skipping tests spread over many builds, not only on those that get fully skipped.
Our observation implies that skipping full builds was a better strategy for saving cost.

\HIDDEN
{
}

\mysubsection{Comparing Techniques}
We first observed that \obs{SB\_Mach19\_C and SB\_Jin20\_C skipped fewer builds than their counterparts that were trained only with data within the same project (SB\_Mach19\_W, SB\_Jin20\_W)}.
After having been trained with a more diverse set of build and tests (across many projects), these techniques became less confident to skip them.
ST\_Herzig15 saved very low ratio of build time despite saving a large ratio of tests.
This is because it very rarely skips tests that failed many times in the past --- regardless of the code changes in the build.
So, within each build, it very rarely skipped the tests with the most past failures --- thus very rarely skipping builds fully.
SB\_Abd19 saved a median 21\% build time, which is a relatively high amount, considering that it only skipped builds with non-executable changes, \eg that only changed formatting or comments.
ST\_Mach19\_W and ST\_Gligoric15 skipped a relatively high ratio of build time (competitively with build selection techniques) because they skipped many full builds.
This is because they analyze the relationship between code changes and tests inside a build.
ST\_Gligoric15 skips all tests that cannot execute the code changes, and ST\_Mach19\_W considers the distance between the changes and the tests in its predictor.
This allows both techniques to fully skip those builds in which no test can execute the code changes --- \ie when only non-executable code was changed, or when no tests exist to execute the changes.
SB\_Jin20\_W and SB\_Jin20\_S saved high ratios of build time, since they both focused on skipping full builds.
While SB\_Jin20\_S provided higher savings, we expect it to also skip a higher ratio of skipped failing builds (see \cref{sec:d2}) --- SB\_Jin20\_S simply skips builds with $<$4 commits.
Finally, SB\_Hassan17\_W and SB\_Hassan17\_C skipped too much build time (higher than the perfect baseline).
This is because they mostly rely on the status of the previous build, which is unknown if skipped.
So, as soon as they observe a passing build, they recurrently skip all subsequent builds.

\begin{figure*}%
		\centering
		\subfloat{{\includegraphics[width=0.4\linewidth]{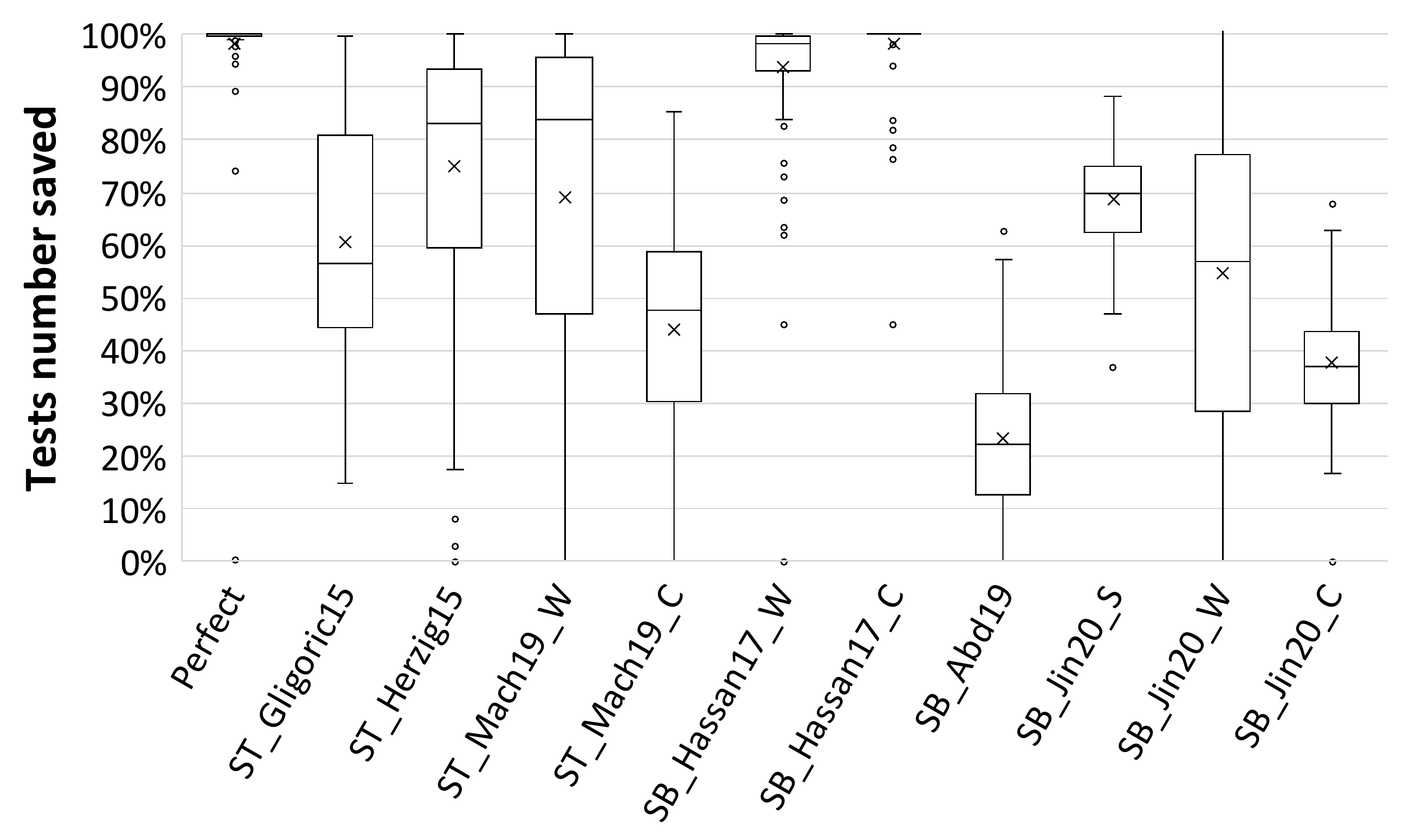} }}%
		\qquad
		\subfloat{{\includegraphics[width=0.4\linewidth]{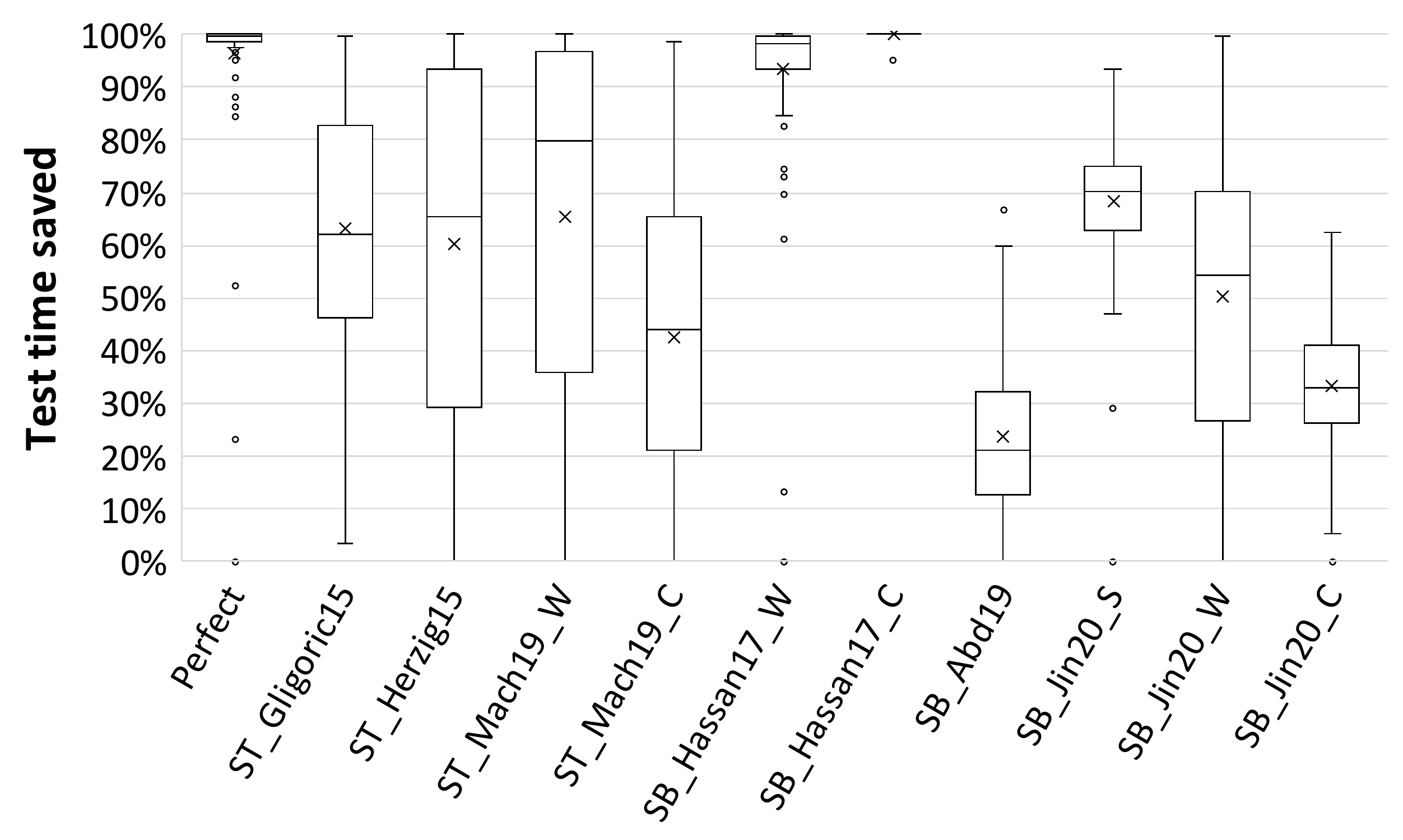} }}%

		\centering
		\subfloat{{\includegraphics[width=0.4\linewidth]{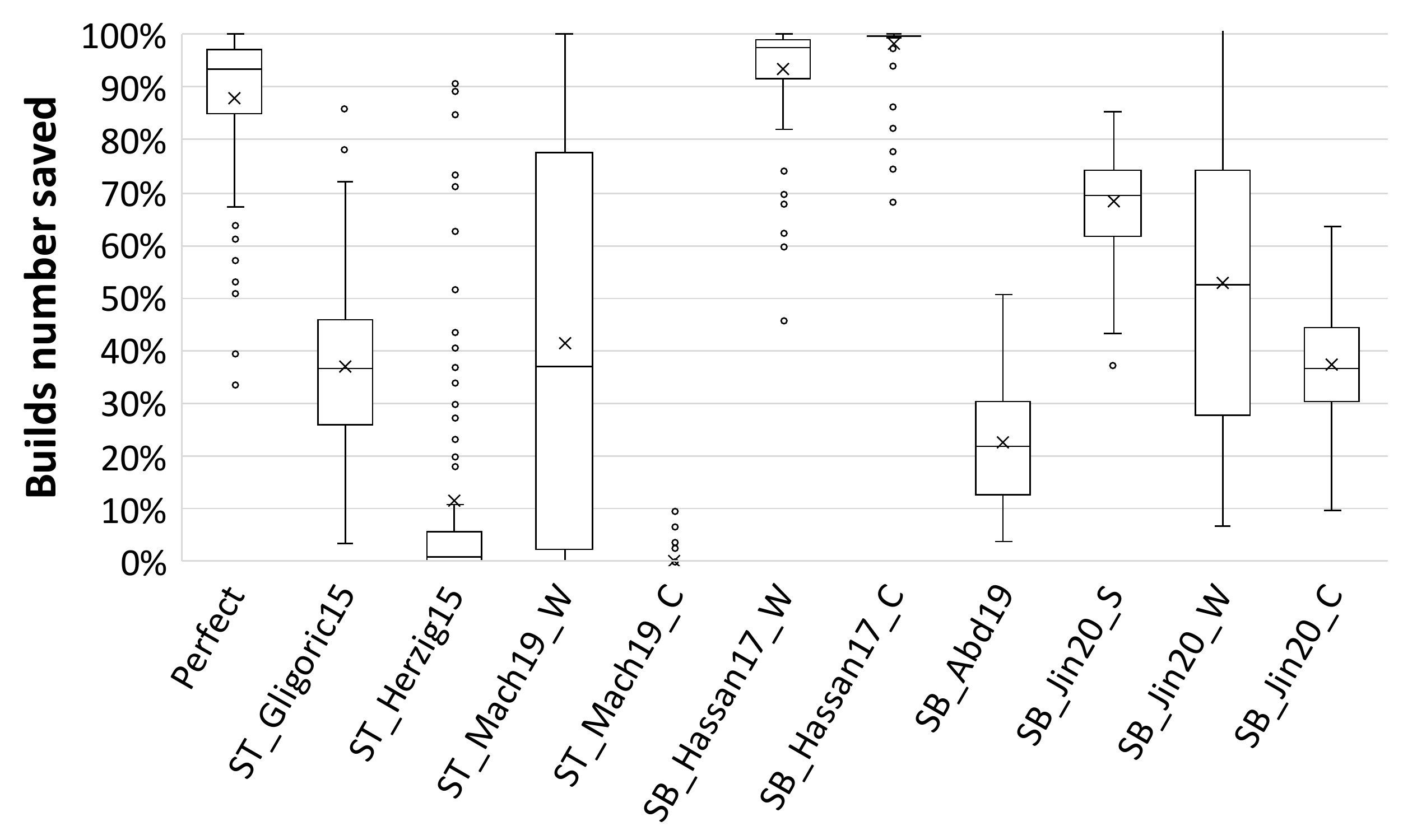} }}%
		\qquad
		\subfloat{{\includegraphics[width=0.4\linewidth]{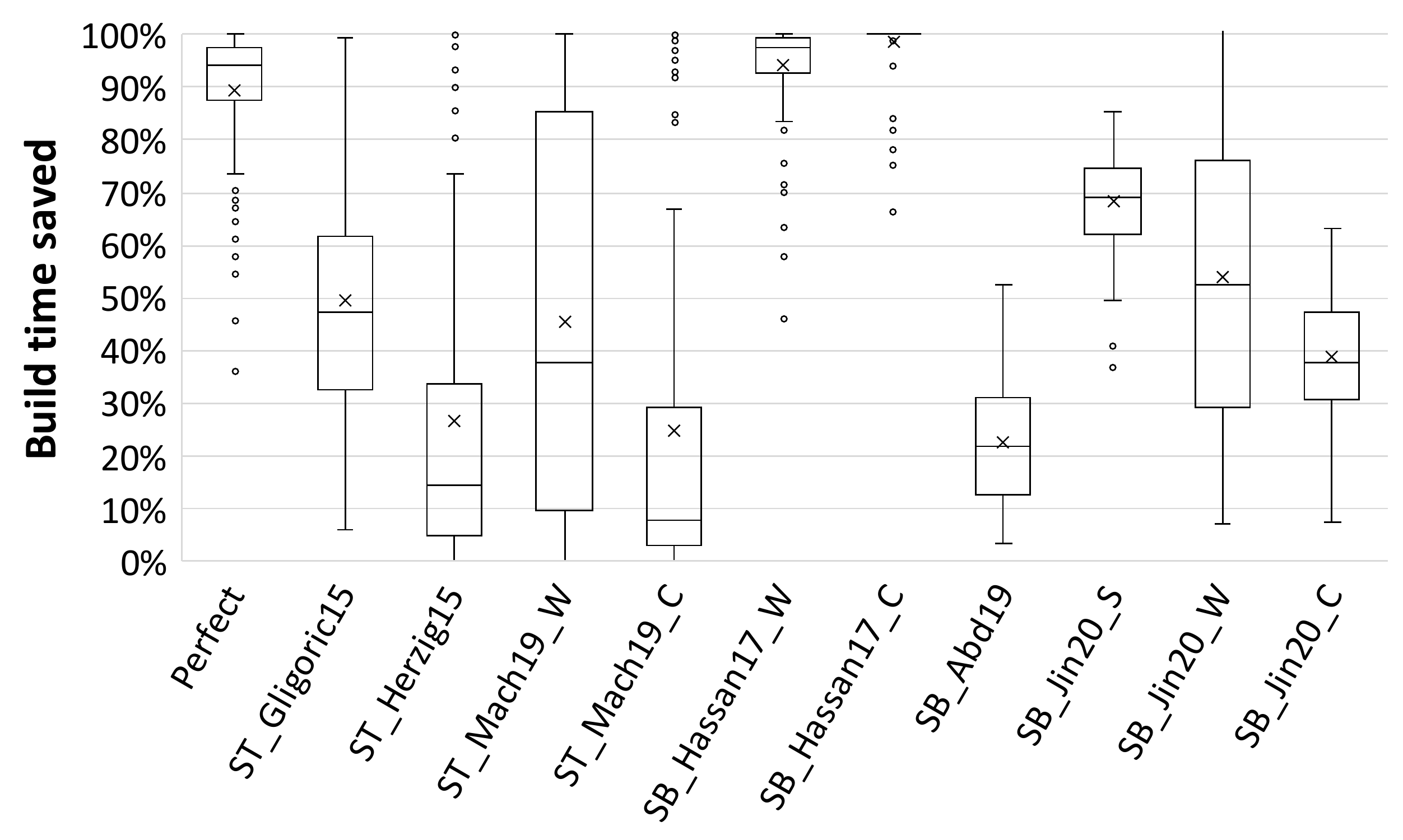} }}%
		\caption{Results for Cost Saving Metrics. Prioritization techniques not included, since they do not skip tests/builds.}
		\vspace{-.1in}
		\label{fig:amount_save}%
\end{figure*}

\begin{figure}%
		\centering
		\subfloat{{\includegraphics[width=0.8\linewidth]{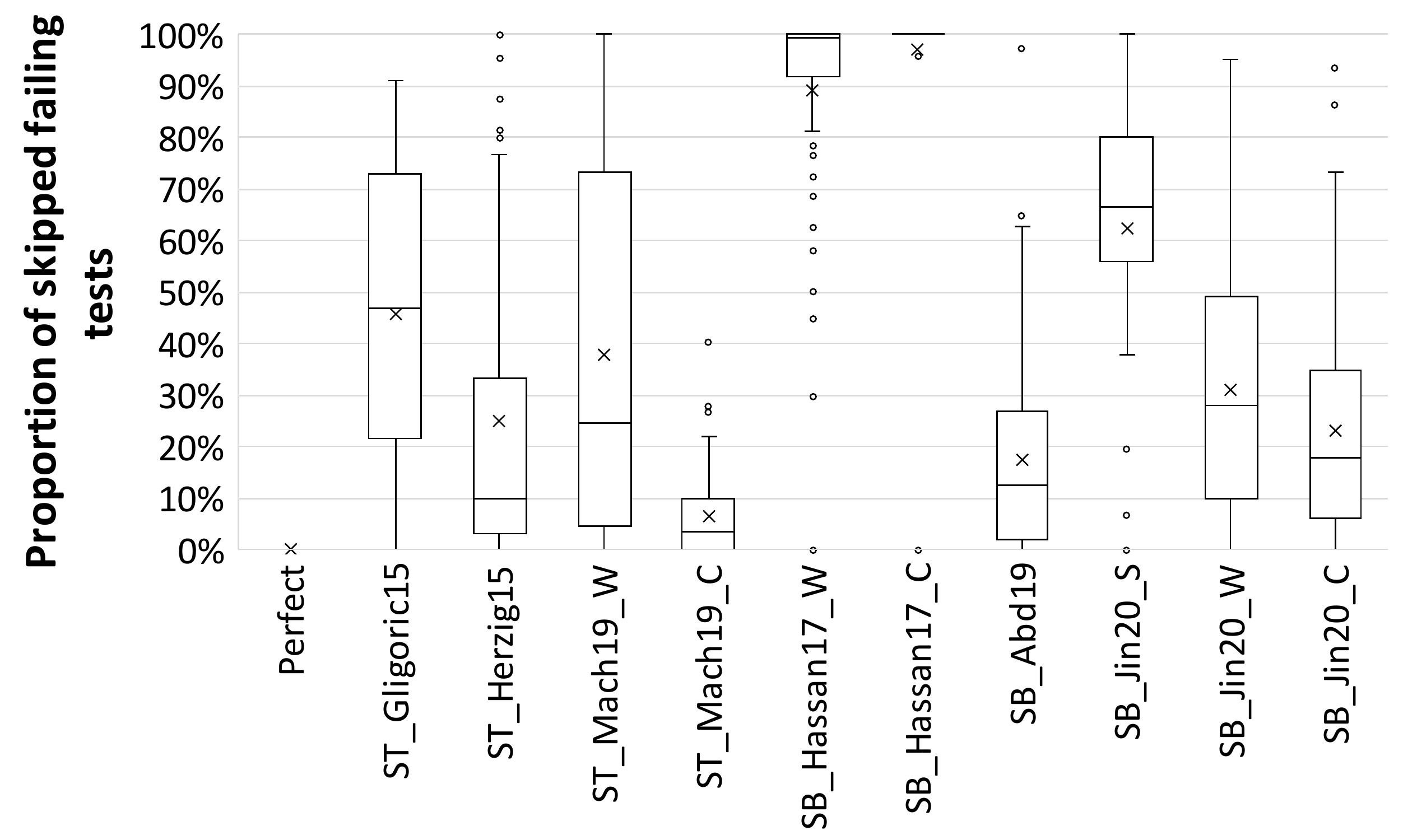} }}%
		\qquad
		\subfloat{{\includegraphics[width=0.8\linewidth]{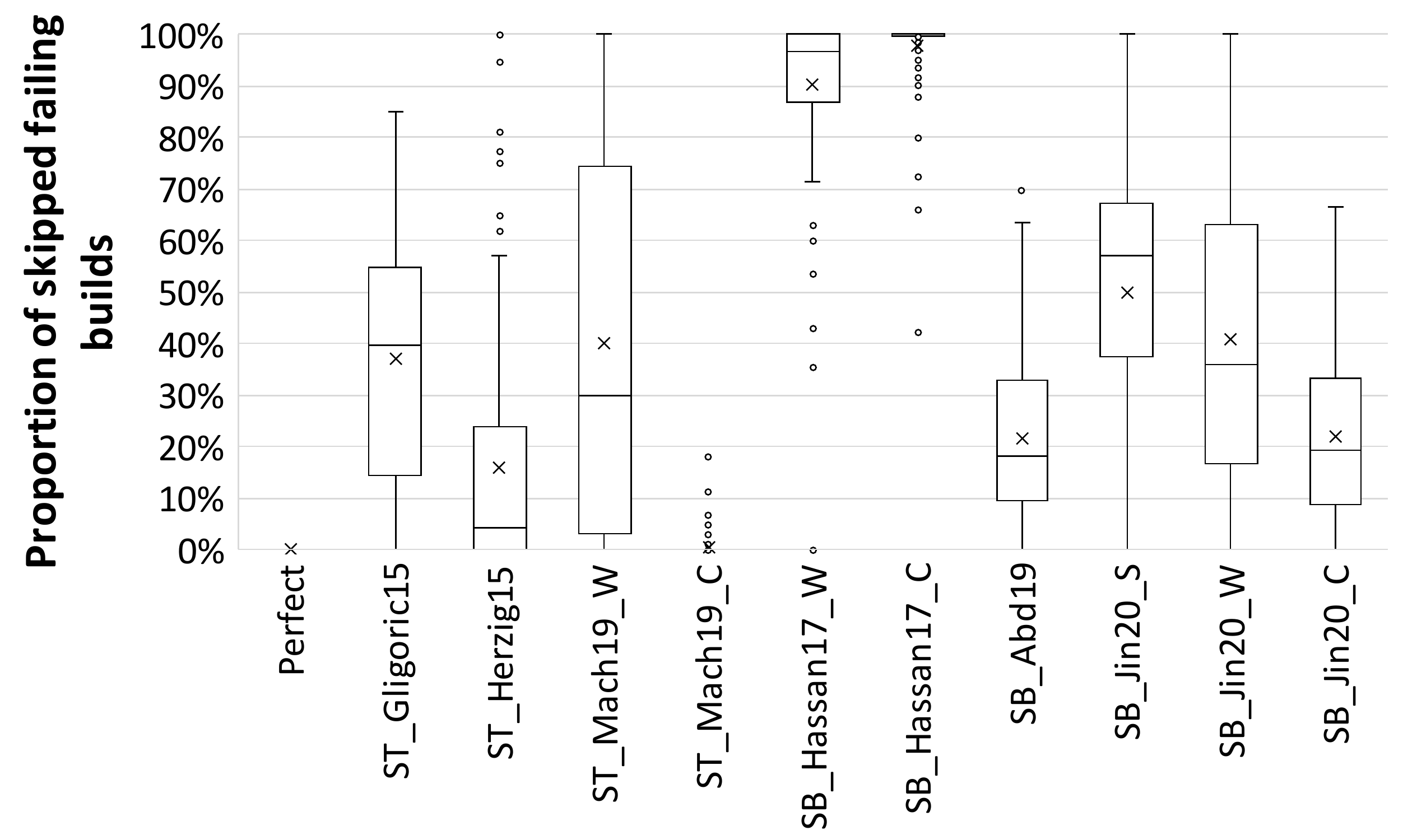} }}%
		\vspace{-.1in}
		\caption{Results for Missed Failure Observation Metrics. Prioritization techniques not included, since they do not skip tests/builds.}
		\vspace{-.2in}
		\label{fig:skip_proportion}%
\end{figure}

\subsection{D2: Missed Failure Observation}
\label{sec:d2}


\subsubsection{Studied Metrics}

\mysubsection{Proportion of skipped failing tests}
This metric measures the undesired side effect of cost-saving techniques skipping some of the failing test cases.
It was used by ST\_Herzig15 \cite{herzig2015art}.

\mysubsection{Proportion of skipped failing builds}
This metric measures the proportion of failing builds that are skipped among all failing builds.
It was covered in SB\_Jin20 \cite{jin2020_icse}.


\subsubsection{Analysis of Results}

\mysubsection{Comparing Metrics}
All techniques generally skipped a very similar ratio of failing tests than builds, with small differences.

\obs{ST\_Mach19\_C, ST\_Herzig15, ST\_Gligoric15, SB\_Jin20\_S skipped a slightly higher ratio of failing tests than builds.}
This is explained by test-granularity techniques skipping partial builds in addition to full builds, and thus they also skipped a higher ratio of failing tests.
The case of SB\_Jin20\_S is different: it skipped a higher ratio of tests because it skipped fewer builds with no failing tests --- few changed $<4$ files. 

SB\_Abd19, SB\_Jin20\_C, ST\_Mach19\_W and SB\_Jin20\_W skipped a slightly higher ratio of failing builds than tests.
This means that these techniques skipped failing builds with lower than average (or no) failing tests, \eg failing due to configuration or compilation errors (which amount to 35\% of failing builds).
Finally, SB\_Hassan17\_C and SB\_Hassan17\_W skipped most failing (and passing) tests and builds.

\mysubsection{Comparing Granularities}
\obs{Build-granularity techniques generally skipped higher ratios of failing builds and tests than test-granularity techniques --- except for SB\_Abd19}.
They generally skipped a higher ratio of all tests and builds.


\HIDDEN
{





}

\mysubsection{Comparing Techniques}
If we rank techniques on these two metrics of side-effect, we observe that they rank almost exactly in the opposite order as they would according to build time saved (for D1).
This shows a clear trade-off between cost-saving and its side effect of skipping failures.

\begin{figure*}%
		\centering
		\subfloat{{\includegraphics[width=0.4\linewidth]{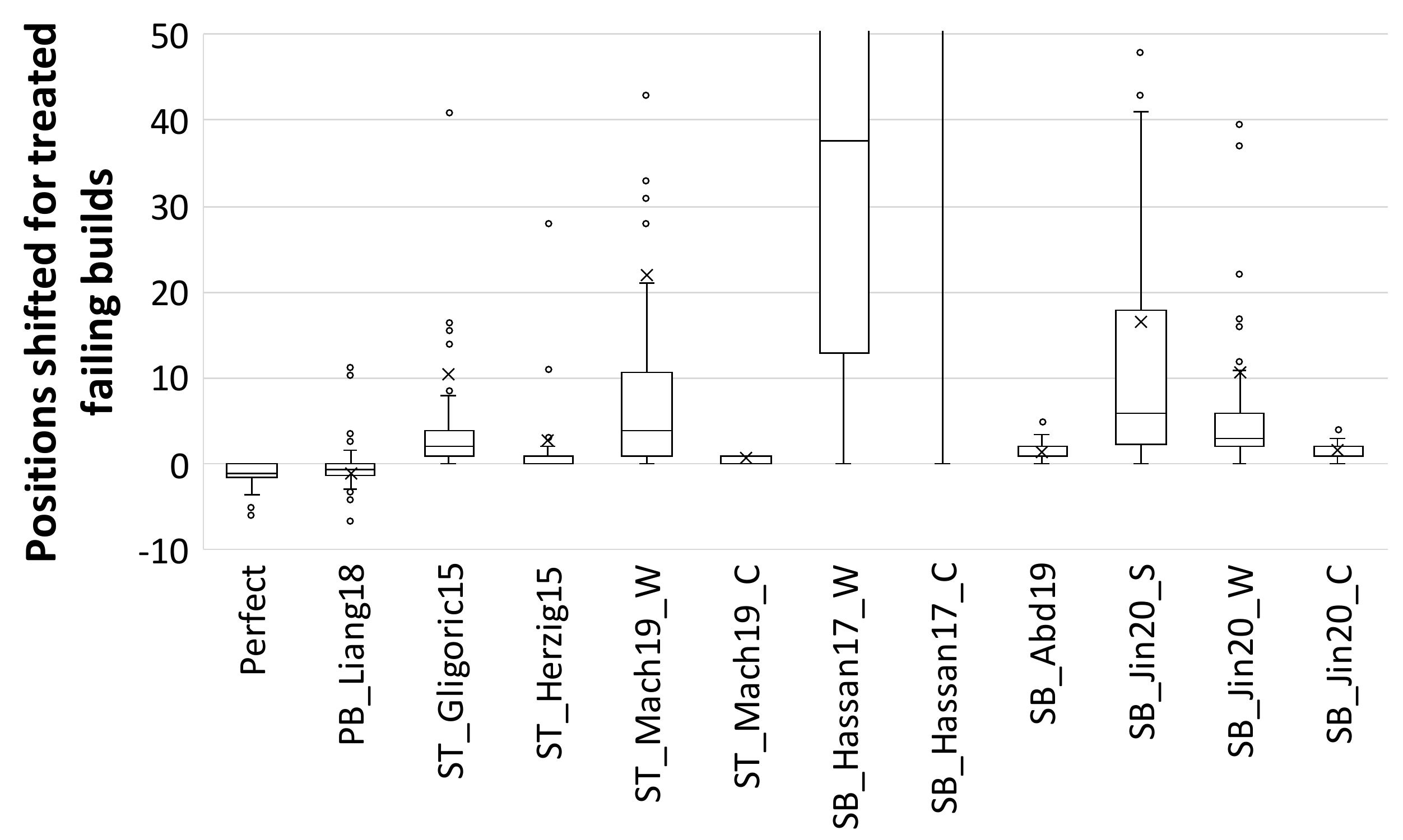} }}%
		\qquad
		\subfloat{{\includegraphics[width=0.4\linewidth]{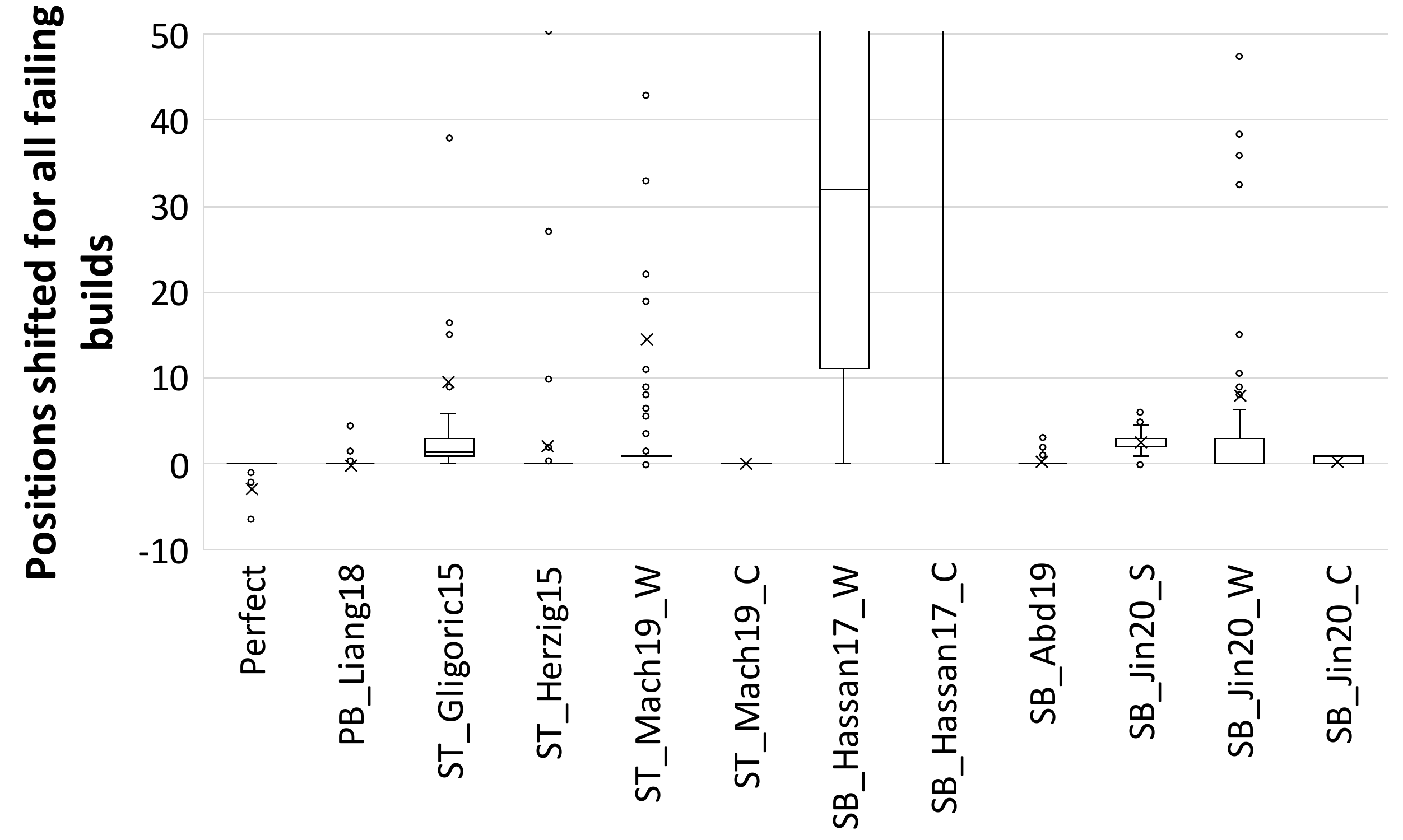} }}%


		\centering
		\subfloat{{\includegraphics[width=0.4\linewidth]{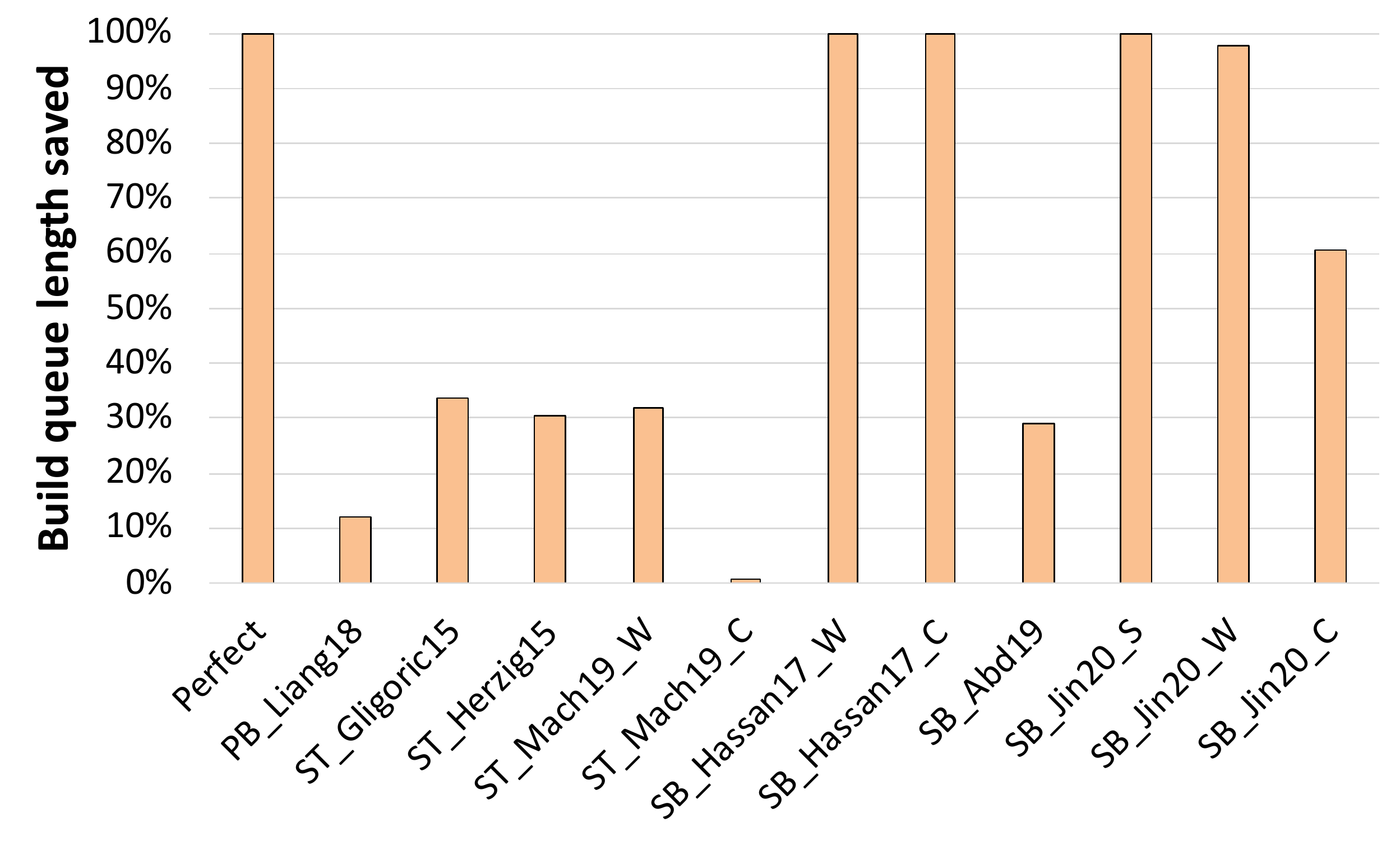} }}%
		\qquad
		\subfloat{{\includegraphics[width=0.4\linewidth]{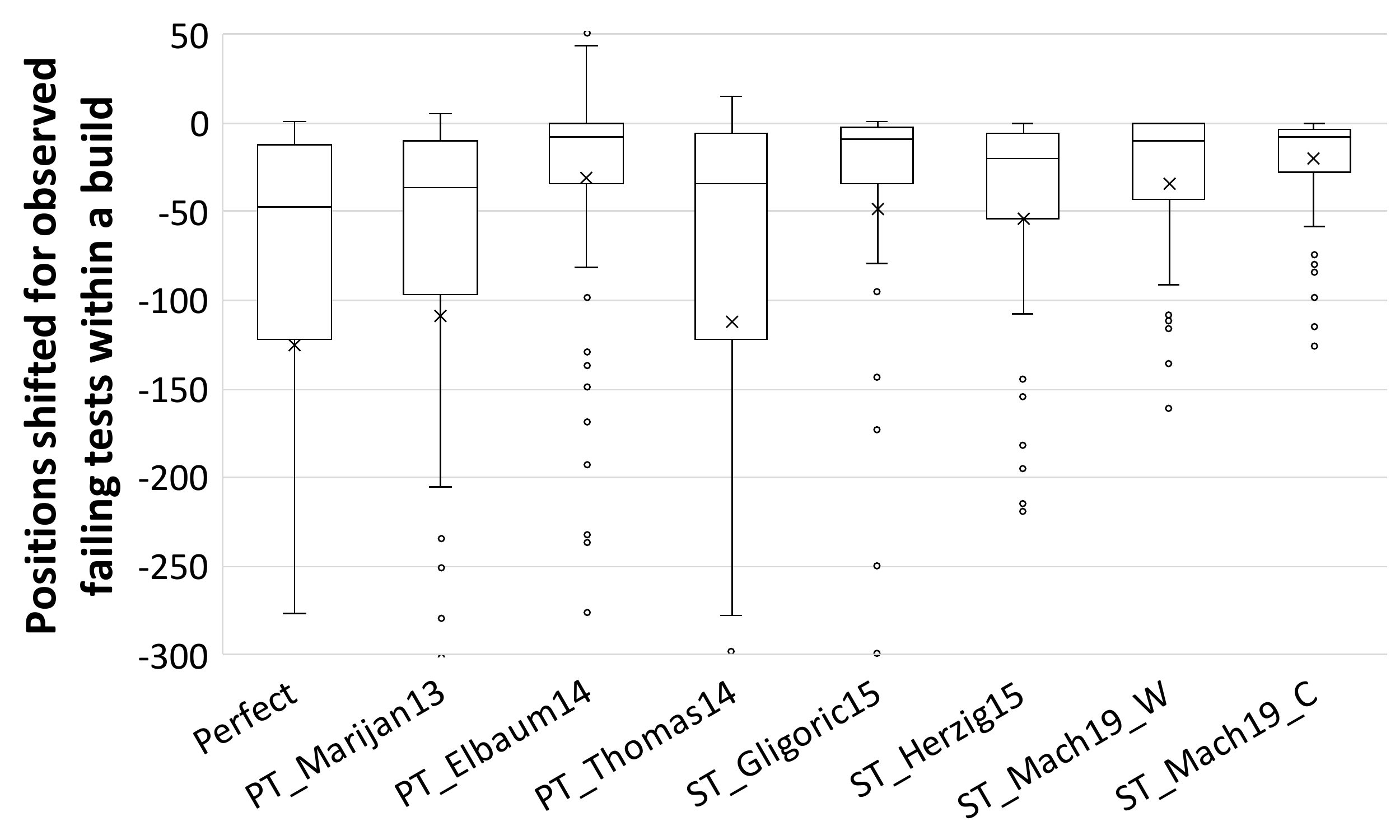} }}%
		\vspace{-.1in}
		\caption{Results for Time-to-feedback Reduction Metrics.}
		\label{fig:position}%
		\vspace{-.2in}
\end{figure*}







\section{Empirical Study 2. D3: Time-to-feedback Reduction}
In D3, we study how much prioritization techniques advance the observation of failures and how much the side effect in D2 will influence it.
So, we study all the time-to-feedback and computational-cost reduction techniques.


\subsection{Studied Techniques}
\label{sec:tech2}
We only describe here the techniques that we did not describe in earlier sections: prioritization techniques.

\subsubsection{Test-prioritization Techniques}
For this family of techniques, we replicated all the test-prioritization techniques that were proposed for improving CI: PT\_Elbaum14 \cite{elbaum2014techniques} and PT\_Marijan13 \cite{marijan2013test}.
To further extend this study, we also replicated the state-of-the-art test case prioritization (TCP) technique.
We chose the technique that provided the highest effectiveness in the most recent evaluation of TCP techniques \cite{luo2018assessing}: PT\_Thomas14 \cite{thomas2014static}.
TCP was a rich research area before CI became a common practice, \eg \cite{mostafa2017perfranker, thomas2014static, elbaum2002test, rothermel2001prioritizing}.
We apply these techniques to prioritize tests within each build. 

\noindent\textbf{PT\_Marijan13 \cite{marijan2013test}} prioritizes tests that failed recently or have a shorter duration.
Tests are ordered based on their historical failure data, test execution time and domain-specific heuristics.

\noindent\textbf{PT\_Elbaum14 \cite{elbaum2014techniques}} favors tests that failed either recently or a long time ago.

\noindent\textbf{PT\_Thomas14 \cite{thomas2014static}} uses topic modeling to diversity the tests that get executed earlier.
Every prioritized test is selected if it contains the most different topics from the previous test in its identifiers and comments.
The rationale behind this is that similar tests often find similar problems.



\subsubsection{Build-Prioritization Techniques}
To the extent of our knowledge, only one technique has been proposed to prioritize software builds, PB\_Liang18 \cite{liang2018redefining}.
\noindent\textbf{PB\_Liang18 \cite{liang2018redefining}} executes builds containing a recently-failing and recently-non-executed test in a collision queue.
We apply PB\_Liang18 to prioritize builds within a build waiting queue, as its previous evaluation did \cite{liang2018redefining}.
Queues form when build executions overlap in time.

\subsection{Studied Metrics}

\subsubsection{Positions shifted for observed failing tests within a build}
measures the shifted positions for all observed failing tests (prioritized or not).
A similar metric to this one was used in the evaluations of PT\_Marijan13 \cite{marijan2013test}, PT\_Elbaum14 \cite{elbaum2014techniques}, and PT\_Thomas14 \cite{thomas2014static}.
For test-selection techniques, we measure the average number of shifted positions for all remaining tests --- when a test is skipped, the next one can now run one position earlier.

\subsubsection{Positions shifted for treated failing builds}
measures the number of builds between every treated (delayed/advanced) failing build's original observation position and its new position. 
This metric was studied by SB\_Jin20 \cite{jin2020_icse}.
For test-granularity techniques, this metric is not impacted, since the build is still executed in the same position.
For build-selection techniques, we consider that when a build is skipped, it will run as the next build (its tests will run on it).

\subsubsection{Positions shifted for all failing builds}
measures the same as the previous one, but now across all failing builds.
PB\_Liang18 used a similar metric in its evaluation \cite{liang2018redefining}.
Through this metric, we can understand the impact of 
the previous metric over all builds.

\subsubsection{Build-queue-length saved} 
This is a metric designed by us to measure how applying a technique could relieve the collision problem: when multiple builds are waiting to be executed within a limited resource.
We follow the same configuration in PB\_Liang18's paper.
The build-queue-length refers to the median number of builds waiting ahead for each build in each project.
With a pre-experiment on all projects, we find that for only one project - "Rails/Rails", the median value of every build's waiting queue is bigger than 0.
Thus, we only report the result for this metric on that project.

\subsection{Analysis of Results}

\mysubsection{Comparing Metrics}
When comparing positions shifted for
treated failing builds vs. all failing builds,
\obs{for all techniques, the advance (PB\_Liang18) or delay (others) that they introduce in the observation of failing builds is much lower when measured across the whole population of failing builds.}
The upside of this is that the undesired effect of most techniques (\ie delay of failure observation) is very low across all failing builds (median 0--2 builds).
The downside is that the desired effect of PB\_Liang18 (\ie advance of failure observation) is also very low across all failing builds (median 0 builds).

Next, we compare the performance of test selection techniques (\ie the only overlapping technique family)
in
the positions that observed failing tests shifted within a build vs.
the positions that failing builds shifted across all builds.
We observe that \obs{test selection techniques provided some advancement in the observation of test failures (lower than most test prioritization techniques), while introducing a very low delay in observation of build failures (median 0--2)}.


\mysubsection{Comparing Granularities}
We did not observe a substantial difference when comparing granularities --- we observed stronger differences when comparing techniques.


\mysubsection{Comparing Technique Strategies}
When comparing technique strategies (prioritization vs. selection), 
test-selection techniques provided some advancement in the observation of failing tests within a build,
but test-prioritization techniques provided better results overall (except PT\_Elbaum14).

\mysubsection{Comparing Techniques}
PT\_Marijan3 and PT\_Thomas14 behave very similarly --- despite their different approaches to prioritization --- and they are both close to perfect, prioritizing most tests correctly.
PT\_Elbaum14 provides a lower advancement of test failures (also lower than many test-selection techniques), since it uses a simpler criterion --- prioritizing tests that were executed very recently or a long time ago.
All test-selection techniques provided a very similar advancement of test-failure observation, except ST\_Herzig15 which was slightly better.
Interestingly, ST\_Herzig15 was one of the techniques with the lowest delay in build-failure observation (median 0 for all failing builds).
At the build-granularity, PB\_Liang18 had a very low impact in prioritizing builds because builds very rarely occurred concurrently in our dataset --- only the Rails project had a meaningful number of concurrent builds.
An important metric in PB\_Liang18's original evaluation was the savings in the build-queue length.
We plot the results for all techniques for this metric in \cref{fig:position}.
Interestingly, we also observed that test-selection and build-selection techniques also had a strong impact in this metric --- less so for test-selection techniques and SB\_Abd19 because they skip fewer full builds (see \cref{sec:d1rq1}).
Regarding build-selection techniques, those that saved more builds (see \cref{sec:d1rq1}) also saved more in the build-queue-length metric, but also introduced higher delays in build-failure observation.

\HIDDEN{
	Discussion:
	Why is this surprising? We didn't know if test-granularity techniques...
	... saved more or fewer full builds than build-granularity techniques
	... saved many or few partial builds than build-granularity techniques

	Also, while it is obvious that we could combine test skipping and build skipping, we don't know if it would be counter-productive in terms of missing failures.

	Can we really save more cost targeting slower builds?
	This opportunity, however, only applies if build duration within a project varies over time.
		Can we give a number?

	Motivation
	This field is becoming extremely popular - in academia and industry
	Somebody has to design how to move it forward

	Put all findings together, and provide recommendations

	Some of our findings may be obvious in retrospect, but:
	- no technique explicitly mentions the opportunities that we highlight
		- so they are only obvious in retrospect
	- we didn't know if techniques were achieving our highlighted opportunities already
		- there may not have been much space for improvement
	- we are setting the direction to move this field forward
		- we are the first to explicitly propose our opportunities/implications for improvement, and to provide empirical evidence for them
	- It's not obvious what would work and what wouldn't
		- it clearly was not obvious to the techniques' original authors!!

	Could we guess the results based on a technical analysis of the techniques?
		No, because we don't know the shape of the data in any of the dimensions that matter!
		We could have studied separately how common each desirable characteristic is
		But it's better to simply run the techniques.

	We are also contributing an evaluation framework, with metrics to allow evaluations across-the-field
		what's the closest published thing in this respect?

}


\section{Answers for Research Questions and Implications}
\label{sec:discussion}
We synthesize our observations and we lay out their implications to advance this area of research.

\subsection{D1: Computational-cost Reduction}
\label{sec:discussion-d1}

\subsubsection{RQ1: What design decisions did not help?}
First, we report on \textbf{missed opportunities} for saving more computational cost.
Cost-saving techniques focused on skipping passing builds and tests, but they \textbf{did not specifically target those that would provide the highest savings}, \ie slower tests, slower builds, or all tests in a build (in the case of tests-selection).
This is demonstrated by the fact that build-granularity techniques saved similar ratios of test number, test time, build number, and build time; and that test-granularity techniques saved similar ratios of test number and test time, and lower ratios of build time than test time.

We also learned that \textbf{training cost-saving techniques across projects} harmed their predictions.
In other fields, training with data from multiple projects is considered to increase the accuracy of predictors.
For cost-saving techniques, though, this exposed the techniques to more diverse sets of failures, making more builds/tests ``look like a failure'', resulting on the predictors saving less cost (being less inclined to skip builds and tests).

Test-selection techniques were also limited in the cost that they could save when \textbf{they did not target saving full builds} --- ST\_Mach19\_C and ST\_Herzig15 saved very low build time despite saving a high ratio of tests.
An additional aspect that contributed to ST\_Herzig15 saving limited build time (despite saving high number of tests) is that \textbf{it only used features characterizing the tests}, but not the code changes in the build --- \eg missing the opportunity to skip full builds for no-code changes.

\subsubsection{RQ2: What design decisions helped?}
Other design decisions allowed techniques to save high cost.
A particularly useful design decision was \textbf{trying to predict seemingly-safe builds and tests} --- SB\_Abd19 saved 21\% builds simply by skipping builds with no-code changes, and ST\_Gligoric15 saved 36\% builds skipping tests that did not cover the code changed in the build.

Another decision that provided high cost savings was to \textbf{skip full builds instead of individual tests} --- thus also saving build-preparation time.
Skipping all tests in a build allows to skip the time to prepare the build (\ie compilation and other overhead like virtual machine preparation), and we observed that \textbf{build-preparation takes a large portion of build time}.
An illustrative example is how ST\_Gligoric15 and ST\_Herzig15 saved about the same ratio of test time, but ST\_Gligoric15 saved much higher build time because it saved a much higher ratio of full builds.

Test-selection techniques, however, performed really well in terms of saving a high ratio of tests (84\% by ST\_Herzig15 and 80\% by ST\_Machalica\_W).
This is because they could save some cost spread out across many builds --- \ie \textbf{skipping partial builds achieved high cost savings}.
However, the test-selection \textbf{techniques that skipped full builds also achieved high savings}.
Intentionally or not, ST\_Gligoric15 saved many full builds by simply skipping all tests that did not cover the changed code.
ST\_Mach19\_W also saved many full builds by approximating the same idea: one of its predictor's features is the distance between the changed code and the test.

\subsubsection{Implications for Future Techniques}
Our results have multiple implications for the design of future techniques.
First, we encourage future techniques to consider \textbf{hybrid approaches} to save both full builds and also partial builds, \ie to save cost at both build and test granularity.
Future techniques should also leverage the beneficial factors that we already observed, such as \textbf{skipping full builds with no-code changes or no tests to cover them}.
To save more full builds, novel prediction features could be designed, \textbf{targeting slower builds} if possible --- which no existing technique attempts.
To save more tests, existing techniques already provide very useful features (saving a high ratio of tests), but other new features could be designed to target \textbf{saving more and slower tests}, and considering the \textbf{relationships between the tests and the code changes} in the build.
Finally, our observations also show that \textbf{build time saved} is the metric that most comprehensively shows the cost saved by all existing techniques --- even though cross-referencing multiple metrics allows for additional observations, as we did in this study.

\subsection{D2: Missed Failure Observation}
\label{sec:discussion-d2}

\subsubsection{RQ1: What design decisions did not help?}
In terms of the proportion of builds and tests that were skipped by cost-saving techniques, we generally observe that \textbf{the decisions that made techniques save higher cost also made them make more mistakes}, \ie skip higher ratios of failing builds and tests.
It was also particularly interesting that \textbf{seemingly-safe techniques} --- SB\_Abd19 and ST\_Gligoric15 --- still \textbf{showed pretty high ratios of skipped failing builds and tests}.
Our study thus shows that skipping builds with no-code changes or without tests to execute them is not enough to guarantee that they will not fail.
A quick look discovered that the builds and tests skipped by these techniques failed for different reasons, such as configuration or compilation errors (present in 35\% of failing builds).

\subsubsection{RQ2: What design decisions helped?}
One design decision that reduced the skipped failing tests and builds was \textbf{training techniques across projects}.
All the \_C variants skipped lower ratios than their \_W counterparts (except SB\_Hassan17\_C).
Also \textbf{test-granularity techniques generally skipped lower ratios of failing tests} than build-granularity techniques did of builds.

\subsubsection{Implications for Future Techniques}
These results imply multiple recommendations for future techniques.
First, future techniques should design \textbf{novel features to predict failures that are caused by no-code changes}, \eg configuration changes, to avoid assuming that seemingly-safe builds will not fail.
Second, future techniques should attempt to \textbf{break this trade-off between saving cost and skipping failures}.
Existing techniques generally increase cost savings by also increasing missed failure observations.
Future techniques should attempt to improve one of the two dimensions by keeping the other one fixed (or optimal).
Finally, future studies should propose \textbf{new metrics to better assess the trade-off between cost-saving and skipped-failures} of various techniques --- since most techniques succeed in one at the expense of the other.
SB\_Jin20 \cite{jin2020_icse} proposed the harmonic mean of the two as a balanced metric, but further study is granted to understand whether both should be valued equally or in a weighted manner --- particularly considering the much higher ratio of passes to failures in CI datasets.

\subsection{D3: Time-to-feedback Reduction}
\label{sec:discussion-d3}

\subsubsection{RQ1: What design decisions did not help?}
Unsurprisingly, \textbf{build-selection techniques did not advance the observation of build failures at all}, but at least they introduced very low delays in the observation of failing builds (and also saved some computational cost).
Similarly, \textbf{test-selection techniques also introduced a small delay in the observation of test failures}.
\textbf{Build-prioritization also showed very limited advancement in observing failing builds}, but that was mainly because only one of our studied projects (open-source) had some contention in the build queue.
We expect that industrial software project would obtain a much higher benefit from this approach.
Finally, we also observed that the build-selection techniques that produced \textbf{higher cost savings also introduced higher delays in build-failure observation}, showing again the tension between both goals.

\subsubsection{RQ2: What design decisions helped?}
\textbf{The best techniques to provide early feedback were test-prioritization techniques}.
In fact, PT\_Thomas14 provided near perfect results.
We also found that \textbf{test-selection techniques provided lower, but competitive advancement of test failure observation}, while also providing some cost savings.
For example, ST\_Herzig15 provided high advancement of test-failure observation within a build, with very low delay of build-failure observation, while also saving some computational cost.
Similarly, we observed that \textbf{build-selection techniques could also provide reductions in build-queue-length that were competitive with build prioritization}.

\subsubsection{Implications for Future Techniques}
For future techniques, we recommend to \textbf{combine test prioritization with test selection techniques} --- since prioritization techniques could stop after the first failure is identified, and save the cost of running the remaining tests.
We found that test-prioritization techniques already reached very high results (PT\_Thomas14 is near perfect), so the features that they use could be also very useful for test selection to save cost.
Conversely, existing test-selection techniques that already perform very well for cost-savings (\eg ST\_Herzig15) could be improved in their ability to advance failure observation.
Similarly, we recommend to further study the application of \textbf{build-selection techniques to provide early observation of build failures} by reducing the build queue via skipping builds in industrial projects in which parallel build requests are a larger issue.
Finally, there is also space to develop new metrics that could capture the balance that techniques provide across all dimensions D1--D3.

\subsection{Standing on the Shoulders of Giants}
\label{sec:extended-observations}

Our findings confirm and extend previous work:

\subsubsection{D1}
Beller \etal \cite{beller2017oops} observed that test time is a low proportion of build time. 
We extend this observation by finding that our studied test-selection techniques infrequently skipped full (all tests within) builds, which strongly limited their cost-saving ability. 
We thus recommend test-selection to incentivize skipping full builds to save higher cost in CI.

\subsubsection{D2}
Jin and Servant \cite{jin2020_icse} observed a trade-off of higher cost savings incurring more missed build failures in their technique. 
We extend this observation by finding that all our studied techniques were affected by that trade-off (techniques ranked equally by cost savings as by missed failures). 
We additionally identified clear strategies that made techniques miss fewer failures: training across projects, and operating at test granularity. 
We also observed that a seemingly-safe technique \cite{abdalkareem2019commits} still missed a high ratio of failures. 
Finally, we elicited the need for better prediction of safe builds, and new metrics to compare trade-offs.

\subsubsection{D3}
Herzig \etal \cite{herzig2015art} found that their test-granularity technique incurs low delay in build-failure observations. 
We extend this observation by finding that all our other studied test-granularity techniques also incur low build-failure-observation delay, measured across all failing builds.

\HIDDEN
{
\section{Implications}

\mysubsection{For researchers} 
Our work provides a set of metrics for the three dimensions when evaluating performances of computational-cost reduction and early feedback time.
Among those metrics, we find that build time seems to be the best metric for evaluating cost reduction, even for test-based techniques.
Other metrics are useful in addition, to observe specific effects, \eg test time can provide detailed information for techniques' preference on test duration.
For evaluating side effects, we find that both metrics have some advantages.
In that case, a new metric should be defined to be able to make both kinds of observations. 
It should measure the proportion of skipped failures by combing skipped builds and tests.
We also find metrics not comprehensive enough when evaluating the dimension of early feedback time.
There should be new metrics coming out, measuring cumulative delay (or advance) for early-feedback techniques, and new metric of balance cost-saving with cumulative delay (for cost-saving techniques)

We also find some opportunities for improving existing techniques.
Regarding to cost-saving, future techniques may want to bridge the gap between test and build selection. A hybrid technique could take advantage of both.
Also, new techniques should aim to: consider skipping builds partially (like test-based techniques), but also consider build-specific factors (like predicting errors not revealed by test failures).
Besides, new techniques could focus on either: best balance, best fully-safe technique, or best savings with lowest delay.
Finally, there is a high potential to improve the safe-skip build selection approaches.

We also make some interesting observations that could guide future research.
We find that in our data set, 35\% of failing builds have no failing test.
This value is similar with previous work's \cite{beller2017oops} results (30\% for Java and 33\% for Ruby).
However, both test and build selection techniques have no specific preference in this kind of builds.
Maybe future techniques should split these builds from the population and analyse them separately.
Besides, we observe that test-selection techniques do have some builds with no test got selected, but they save less proportion of build time compared with the proportion of saved tests.
This is because the build is made up of several phases and test execution belong to one of them, taking a medium proportion (about 40\%).
The phase such as installation or release could also take a lot of time and it becomes a blank space for test selection techniques in CI context.
As a result, there is a big potential to save build time.


\mysubsection{For practitioners}
Our findings can help practitioners decide which technique is best for them to maximize the benefit by understanding the data in their project.
For example, if your project has a lot of collision problems, applying PB\_Liang18 can help you advance feedback time.
Or if your project has redundant builds because of safe modifications such as Readme files, you can probably take advantage of SB\_Abd19 to skip some safe builds.
Or if your project has tests with a strong relationship to your source codes, ST\_Gligoric15 may help save some passing tests and builds since the changes don't touch any of them.



\mysubsection{For tool-builders}
We find that there could be some changes on CI frameworks, \eg Travis CI.
From results of Dimension 3, we find that test-prioritization techniques' performance on the positions advanced for all failing tests within a build is very similar with the performance on the positions advanced for prioritized failing tests within a build.
This reflects that maybe all of the failing tests get prioritized.
We then did an extra experiment and we found that in average more than 95\% of failing tests get prioritized and are not at their original position any more.
This means there is potential to improve the original order of test executions.
In that case, maybe Travis could include test prioritization technique into the build workflow automatically.

}

\section{Threats to Validity}

\subsection{Internal Validity}
To guard internal validity, we carefully tested our evaluation tools on subsets of our dataset while developing them.

Our analysis could also be influenced by incorrect information in our analyzed dataset.
For this, we studied a popular dataset that is prevalent among continuous integration studies: TravisTorrent \cite{beller2017travistorrent}.
Furthermore, many of our studied techniques \cite{abdalkareem2019commits,hassan2017change,jin2020_icse,liang2018redefining} were originally evaluated on TravisTorrent projects.
Additionally, we extensively curated TravisTorrent, removing: toy projects following standard practice \cite{islam2017insights, ni2017cost}, unusable projects for test-granularity techniques, and cancelled builds as in past work \cite{gallaba2018noise,jain2019brief,rebouccas2017does}.
Finally, we also followed the advice in Gallaba \etal's study \cite{gallaba2018noise} to consider the nuance in the TravisTorrent dataset.
We did so in the following ways:
(1) We considered passing builds with ignored failures as passing.
Developers manually flag such failures to be ignored when they cannot officially support them \cite{gallaba2018noise}, and thus should not represent the status of the build.
(2) We considered builds that fail after another failure as correctly labeled, because they flag an unsolved problem, being informative for developers.
(3) We considered failing builds with passing jobs as failing builds. 
If at least one job fails, it signals a problem, informing developers.

Our results may also be affected by flaky tests causing spurious failing builds.
However, CI systems are expected to function even in the presence of flaky tests, since most companies do not consider it economically viable to remove them, e.g., \cite{Machalica2019predictive, micco2017state}.
Besides, cross validation may make unrealistic use of chronological events
To address this problem, we used time-based cross validation \cite{bettenburg2008duplicate}.

Our observed build and test runtimes may have been influenced by the load experienced in the build server at the time.
However, we consider this potential impact to be very low, since we observed that the standard variance in test duration across builds was 0.5 seconds.

\subsection{External Validity}
To increase external validity, we selected the popular dataset TravisTorrent, which has been analyzed by many other research works.
The projects we chose were all Java or Ruby projects, because there are no projects with other programming languages in the data set.
Although these two programming languages are popular, different CI habits in other languages may provide slightly different results to the ones in this study.
Our observations may slightly vary for separate software projects, but our goal was to derive general observations for a real-world population of software projects.


\subsection{Construct Validity}
A threat to construct validity is whether we studied software projects that are similar to those that suffer most accutely from high CI cost and delays in failure observation, e.g., the projects at Google \cite{hilton2016usage} and Microsoft \cite{herzig2015art}.
We studied the TravisTorrent dataset, which is the standard dataset used in the literature to evaluate techniques to save cost in CI \cite{abdalkareem2019commits,jin2020_icse,liang2018redefining,chen2020buildfast}.
One of our studied projects (Rails) is particularly similar to industrial software projects.
Rails was used alongside two other Google datasets to evaluate PB\_Liang18 \etal \cite{liang2018redefining}, and it had similar magnitudes of test suites (thousands), test executions (millions) and test execution time (millions of seconds).

Nevertheless, early observation (or prediction) of build failures is beneficial, regardless of how much load a project's CI system experiences. 
It allows developers to not have to wait for builds to finish, which is the motivation of multiple previous works, e.g., \cite{hassan2017change,abdalkareem2019commits}. 
In particular, Abdalkareem \etal \cite{abdalkareem2019commits} found that developers from small projects --- as small as 168 commits --- also chose to manually skip commits in CI to save time.
These savings can be substantial for the projects in our studied dataset: test-suite runtime varies from project to project (median 2.3 mins, 75th percentile 26 mins) but, more importantly, saving full builds could save much higher cost (median 14 mins, 75th percentile 52 mins). 
Also, many builds (20\%) take longer than 30 minutes \cite{beller2017oops}.
Test-selection could save higher cost if it leaned harder towards skipping full builds, but we found in this study that this incentive is not yet strongly leveraged by our studied test selection techniques.

\section{Related Work}


\subsection{Empirical Studies of CI and its Cost and Benefit}
Multiple researchers focused on understanding the practice of CI, studying both practitioners \eg \cite{hilton2016usage} and software repositories \cite{vasilescu2015quality}.
Vasilescu \etal studied CI as a tool in social coding \cite{vasilescu2014continuous}, and later studied its impact on software quality and productivity \cite{vasilescu2015quality}.
Zhao \etal studied the impact of CI in other development practices, like bug-fixing and testing \cite{zhao2017impact}.
Stahl \etal \cite{staahl2013experienced} and Hilton \etal \cite{hilton2016usage} studied the benefits and costs of using CI, and the trade-offs between them \cite{hilton2017trade}.
Lepannen \etal similarly studied the costs and benefits of continuous delivery \cite{leppanen2015highways}.
Felidr\'e \etal \cite{felidre19} studied the adherence of projects to the original CI rules \cite{fowler2006continuous}.
Other recent studies analyzed testing practices \cite{gautam2017empirical}, difficulties \cite{pinto2017inadequate} and pain points \cite{widder2019conceptual} in CI.

The high cost of running builds is highlighted by many empirical studies as an important problem in CI \cite{hilton2016usage,hilton2017trade,pinto2017inadequate,widder2019conceptual,herzig2015art} --- which reaches millions of dollars in large companies, \eg at Google \cite{hilton2016usage} and Microsoft \cite{herzig2015art}.
People \cite{hilton2017trade,vasilescu2015quality} believe that the benefit of CI is mainly lying in the early fault detection.
Others \cite{hilton2016usage,leppanen2015highways} find that projects adopting CI are able to adopt pull requests and release in a shorter time.
Some also find that CI can help developer team in other areas such as providing a common build environment \cite{hilton2017trade} and increasing team communication \cite{staahl2013experienced}.


\subsection{Approaches to Reduce Time-to-feedback in CI}
A related effort for improving CI aims at speeding up its feedback by prioritizing its tasks.
The most common approach in this direction is to apply test case prioritization (TCP) techniques \eg \cite{luo2018assessing,mostafa2017perfranker,elbaum2014techniques,marijan2013test,elbaum2002test,rothermel2001prioritizing,zhu2018test} so that builds fail faster.
These techniques, even though not designed to work in CI environment, have been claimed to have a potential to provide CI users earlier fault observation.
Another similar approach achieves faster feedback by prioritizing builds instead of tests \cite{liang2018redefining}.
Their paper grants higher priority to those builds that are more likely to fail according to the historical failing information and works well for those projects that have a ton of collision issues.
Naturally, these kinds of techniques don't provide benefit in saving the cost.
In this paper, we study both test-prioritization techniques as well as build-prioritization techniques in terms of advancement of failure observation and compare them with selection techniques.

\subsection{Approaches to Reduce Cost of CI}
A popular effort to reduce the cost of CI focuses on understanding what causes long build durations \eg \cite{ghaleb2019empirical,tufano_icse_2019}.
Thus, most of the approaches that reduce the cost of CI aim at making builds faster by running fewer test cases on each build.
It is found that a ton of passing tests could be saved in this way \cite{labuschagne2017measuring}.
Some approaches use historical test failures to select tests \cite{herzig2015art,elbaum2014techniques}.
Others run tests with a small distance to code changes \cite{memon2017taming} or skip testing unchanged modules \cite{shi2017optimizing}.



Recently, Machalica \etal predicted test case failures using a machine learning classifier \cite{Machalica2019predictive}.
These techniques are based on the broader field of regression test-selection (RTS) \eg \cite{zhu2019framework,zhang2018hybrid,gligoric2015practical,yoo2012regression,yoo2007pareto,rothermel1997safe,rothermel1996analyzing}.
While these techniques focus on making every build cheaper, other work addresses the cost of CI differently: by reducing the total number of builds that get executed.
A related recent technique saves cost in CI by not building when builds only include non-code changes \cite{abdalkareem2019commits,abdalkareem_tse2020}.
They firstly create a rule-based selection technique and then take advantage of machine learning algorithm to improve the accuracy.
Then Jin and Servant propose a build strategy that developing team should skip those less informative passing builds through build outcome prediction.
Finally, other complementary efforts to reduce build duration have targeted speeding up the compilation process \eg \cite{celik2016build} or the initiation of testing machines \eg \cite{gambi2015improving}.
In this paper, we refer cost-reduction techniques as selection techniques.
We pick techniques in both build-selection techniques and test-selection techniques and examines their performance in different cost-saving and fault-observation metrics.

\subsection{Evaluation frameworks for similar techniques}
Multiple research works focus on comparing cross-tool performance with an evaluation framework.
Zhu \etal \cite{zhu2019framework} propose a regression test selection framework to check the output against rules inspired by existing test suites for three techniques.
Leong \etal \cite{leong2019assessing} propose a test selection algorithm evaluation method and evaluate five potential regression test selection algorithms, finding that the test selection problem remains largely open.
Najafi \etal \cite{najafi2019improving} studied the impact of test execution history on test selection and prioritization techniques.
Luo \etal \cite{luo2018assessing} conduct the first empirical study comparing the performance of eight test prioritization techniques applied to both real-world and mutation faults and find that the relative performance of the studied test prioritization techniques on mutants may not strongly correlate with performance on real faults.
Lou \etal \cite{lou2019survey} systematically created a taxonomy of existing works in test-case prioritization, classifying them in: algorithms, criteria, measurements, constraints, scenarios, and empirical studies.

Differently to these works, our study in this paper specifically targets the context of CI, and it has a broader focus than test prioritization or selection.
Our study is the first to compare all the techniques proposed to reduce time-to-feedback or cost in CI, including prioritization and selection techniques, at test and build granularities. 
We performed observations comparing across 2 goals, 3 dimensions, 10 metrics, 2 granularities, and 10 techniques.
Most of our observations required comparisons at broad scope. 
For example: we revealed the need for a new incentive in test selection to skip full test suites (to also save build-preparation time), which would not be relevant in studies outside the scope of CI.

\section{Conclusions and Future work}

In this article, we performed the most exhaustive evaluation of CI-improving techniques to date. 
We evaluated 14 variants of 10 CI-improving approaches from 4 families on 100 real-world projects.
We compared their results across 10 metrics in 3 dimensions.
We derived many observations from this evaluation, which we then synthesized to understand the design decisions that helped each dimension of metrics, as well as those that had a negative impact on it.
Finally, we provide a set of recommendations for future techniques in this research area to take advantage of the factors that we observe were beneficial, and we lay out also future directions to improve on those factors that were not.
We lay out plans to combine approaches at test and build granularities to save further costs, and to combine selection and prioritization approaches to improve on the early observation of failures while also saving some cost.
Such techniques could consider additional history-based prediction features, such as the project's code-change history, \eg \cite{servant2011history,servant2012history,servant2013supporting,servant2013chronos,servant2017fuzzy}, since test-execution history was beneficial for some techniques, \eg \cite{herzig2015art}.
We also discuss the need of future metrics to capture the various characteristics of these techniques in a more holistic way.
In the future, we will work on designing a comprehensive technique that combines selection and prioritization as well as build and test granularities to maximize the benefit of CI while reducing its cost as much as possible.

\section{Replication}
We include a replication package for our paper \cite{xianhao_jin_2020_3696084}.


\bibliographystyle{abbrv}
\bibliography{CI}

\end{document}